\documentclass[%
 reprint,
 amsmath,amssymb,
 aps,
]{revtex4-2}

\usepackage[english]{babel}
\usepackage{graphicx}
\usepackage{dcolumn}
\usepackage{bm}
\usepackage{quantikz}
\usepackage{multirow}
\usepackage{braket}
\usepackage{soul}
\usepackage{hyperref}
\usepackage[label font=bf,labelformat=simple]{subfig}
\usepackage{floatrow}
\usepackage[labelfont=bf,justification=raggedright]{caption}
\usepackage{tabularx}

\begin{document}

\title{First quantum machine learning applications on an on-site room-temperature quantum computer} 

\author{Nils Herrmann}
 \email{nils.herrmann@quantum-brilliance.com}
\author{Daanish Arya}
\author{Florian Preis}
 \email{f.preis@quantum-brilliance.com}
\author{Stefan Prestel}
 \affiliation{Quantum Brilliance GmbH \\ Colorado Tower Industriestr.\ 4 \\ 70565 Stuttgart, Germany}
\author{Mariam Akhtar}
\author{Marcus W.\ Doherty}
\author{Pascal Macha}
\author{Michael L.\ Walker}
 \affiliation{Quantum Brilliance Pty Ltd \\ 60 Mills Road \\ Acton ACT 2601, Australia}

\date{\today}

\begin{abstract}
We demonstrate -- for the first time -- the application of a quantum machine learning (QML) algorithm on 
an on-site room-temperature quantum computer. A two-qubit quantum computer installed at the Pawsey Supercomputing Centre in Perth, 
Australia, is used to solve multi-class classification problems on unseen, i.e.\ untrained, 2D data points. The underlying 1-qubit model 
is based on the data re-uploading framework of the universal quantum classifier \cite{PerezSalinas2020} and was trained on an ideal 
quantum simulator using the Adam optimiser \cite{Kingma2017}. No noise models or device-specific insights were used in the training process.
The optimised model was deployed to the quantum device by means of a single XYX decomposition leading to three parameterised single 
qubit rotations. The results for different classification problems are compared to the optimal results of an ideal simulator. 
The room-temperature quantum computer achieves very high classification accuracies, on par with ideal state vector simulations.
\end{abstract}
\maketitle

\section{Introduction}

\begin{figure*}
    \sidesubfloat[]{\includegraphics[width=0.225\textwidth]{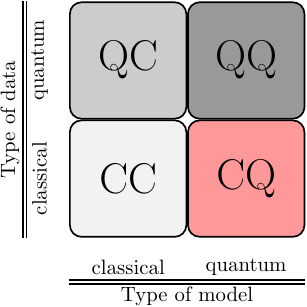}\label{QML_space.fig}}
    \hfill
    \sidesubfloat[]{\includegraphics[width=0.351\textwidth]{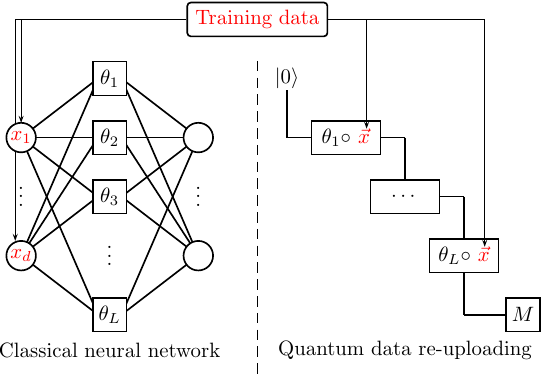}\label{UQC_NN.fig}}
    \hfill
    \sidesubfloat[]{\includegraphics[width=0.288\textwidth]{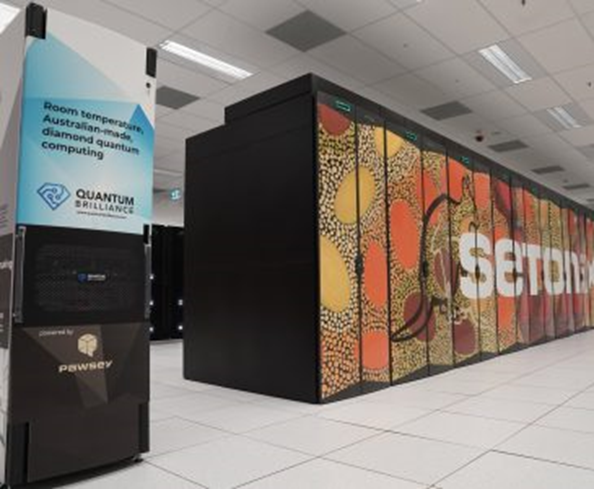}\label{QB_QDK.fig}}
    \caption{\label{KeyFig.fig} \textbf{The first QML application on an on-site room-temperature quantum computer} 
    \textbf{\ref{QML_space.fig}} Following Schuld and Pettruccione \cite{Schuld2021}, the field of quantum machine learning (QML) is separable into 
    four distinct categories. In this work, we focus 
    on classical data processed by a quantum model (``CQ'', highlighted in red). 
    \textbf{\ref{UQC_NN.fig}} The applied model -- the Universal Quantum Classifier (UQC) \cite{PerezSalinas2020} -- is based on the data re-uploading 
    framework \cite{PerezSalinas2020}, which effectively circumvents the no-cloning theorem \cite{Park1970, Wootters1982, Dieks1982}, creating an analogy 
    between single-layered classical neural networks and single qubit quantum circuits.
    \textbf{\ref{QB_QDK.fig}} After having obtained classically trained UQC models, different classification problems were solved using the Quantum Brilliance 
    Quantum Development Kit (QB-QDK) -- an on-site room-temperature quantum computer installed at the Pawsey Supercomputing Centre in Perth, Australia.}
\end{figure*}

Machine learning (ML) is a highly anticipated subfield of artificial intelligence (AI) that aims to find or ``learn'' optimised models 
without explicit instructions \cite{Janiesch2020ML}. The aim is to create models that can effectively predict or classify examples outside the training data in problems where the 
underlying mathematical correlations are either unknown or too complex for conventional approaches. Here, special care needs to be taken to compile training data sets
that ensure a sufficiently general model \cite{Mohri2018Foundations}.   
Especially in light of the newly developed generative ML models, such as the GPT-4 large language model\cite{openai2023}, concerns have been raised addressing 
their increasing training demands both in energy consumption and carbon footprint \cite{Patterson2021, Vries2023}.  
The advent of quantum computing has created high hopes for the field of ML by utilising the exponentially growing 
$n$-qubit Hilbert space as the addressable feature space in ML pipelines. Recently, it has been shown that quantum ML (QML) methods are capable to efficiently 
generalise from fewer training data\cite{Caro2022QMLFew}, can achieve better prediction accuracies\cite{Mishra2021QMLReview}, and possess an inherent advantage 
in handling quantum data\cite{Huang2022QMLAdvantage} compared to their classical competitors, sparking hopes to arrive at a practical quantum advantage -- recently 
coined as Quantum Utility\cite{Herrmann2023}.

Schuld and Petruccione \cite{Schuld2021} separated QML methods 
into four distinct categories as shown in subfigure \ref{QML_space.fig}. 
\cite{Patterson2021, Vries2023}
Here, CC refers to approaches where classical data is processed using classical algorithms. This includes standard classical ML 
approaches but in the context of QML refers to those using only quantum-inspired classical algorithms such as tensor network methods. These have recently 
been used to train ML methods affected by the big data problem in high-energy physics \cite{Felser2021}. The QC category involves QML methods that learn 
from quantum data and apply classical algorithms. Prominent examples include ML-based quantum state tomography \cite{Aaronson2007} and 
classically distinguishing between quantum states \cite{Sasaki2002,Bisio2010}. Of most interest, in the pursuit of Quantum Utility \cite{Herrmann2023}, 
are the CQ and QQ methods, which learn from either conventional or quantum data and train a quantum model, typically a parameterised quantum circuit. 
Tremendous advances in this field have been reported in the literature over the last years. 
For an extensive review of current QML advances, please refer to references \cite{Schuld2015, Biamonte2017, Tychola2023, Zeguendry2023, Melnikov2023}.

In this work, we focus on the CQ sector of QML applications and present -- for the first time -- results of a quantum classification model from a 
quantum computer operating at standard ambient conditions (room temperature and standard pressure) on-site a data centre for high performance 
computing. The model is based on the Universal Quantum Classifier (UQC) \cite{PerezSalinas2020}. 
Very recently, there have been reports of successful applications of the UQC method on ion trap \cite{Dutta2022}, photonic \cite{Ono2023}, and 
superconducting \cite{Tolstobrov2023} architectures. All deployed 
models reach classification accuracies above 90\%, on par or better than classical
competitors with a similar number of trainable parameters \cite{Tapia2022}. The results, reported in this work, differ from the already published applications in two 
fundamental points: 
\begin{itemize}
    \item[(i)] All results in this study were obtained from an on-site nitrogen-vacancy-based spin qubit quantum computer, operating at ambient 
    conditions and mounted on a standard 19" HPC server rack in the Pawsey Supercomputing Centre.
    \item[(ii)] The native performance of the quantum computer was sufficient to achieve high classification accuracies from classically trained (on an ideal 
    simulator using classical hardware) quantum models without the need for additional noise mitigation techniques or on-device training.
\end{itemize}
Therefore, this work marks an important milestone in the pursuit of Quantum Utility \cite{Herrmann2023}. Tackling more complicated and classically difficult 
classification problems using 
larger UQC models with more trainable parameters will ultimately require large quantum processing resources. Here, a simple but robust 
integration to classical computing centres as demonstrated with point (i), will enable massively parallel diamond quantum infrastructures, accelerating the UQC 
training by the parallel distribution of workflows \cite{Nguyen2022}. At the same time, the trained models remain executable on individual diamond quantum processing 
units, which -- due to their ability to operate in ambient conditions -- are edge-deployable. Given a sufficient native device performance, these individual units 
could be used to perform highly complicated inference tasks utilising feature space mappings inaccessible by even the largest classical computing facilities. Following 
a similar train of thought and indicated by point (ii), it might even be possible to achieve Quantum Utility by larger but classically trained UQC models. Here, 
moderately-sized UQC models could be trained on large supercomputing infrastructures and then deployed to individual on-site diamond quantum processing units. 
If the latter are able to outperform classical processing units of similar size, weight, and cost by being faster, more accurate, or requiring less energy 
\cite{Herrmann2023}, a practical quantum advantage could be achieved in the near future.  

The paper is structured as follows: Section \ref{Algorithm} briefly summarises the theory behind the UQC model, 
while section \ref{Application} focuses on the application details covering the classical training process in subsection \ref{Training}, the circuit 
offloading to the quantum processing unit (QPU) in subsection \ref{Deployment} as well as the integration and installation of the quantum device itself 
in subsection \ref{Integration}. Section \ref{Results} highlights the classification results for binary and trinary 
classification problems of two-dimensional classical data, and compares to an ideal state vector quantum simulator.

\section{Algorithmic details}
\label{Algorithm}

The Universal Quantum Classifier \cite{PerezSalinas2020} is based on the data re-uploading framework, effectively 
circumventing the no-cloning theorem \cite{Park1970, Wootters1982, Dieks1982} of quantum computing. This allows the direct 
correspondence between a single-layered classical neural network and a parameterised single qubit quantum circuit shown in Figure \ref{UQC_NN.fig}.

In a single layered classical neural network, the training data $\vec{x}$ is (i) passed on to all connected processing nodes of the hidden layer, 
(ii) processed according to the individual activation functions, and then (iii) re-collected in single or multiple output nodes. The parameters or weights 
$\theta_1,\ldots ,\theta_L$ are optimised with respect to a global cost function. All data connections are applied in parallel, effectively requiring 
a parallel copy operation, which is inherently impossible on quantum computers \cite{Park1970, Wootters1982, Dieks1982}. 
P\'erez-Salinas et al.\ \cite{PerezSalinas2020} proposed to circumvent this problem by serializing all processing nodes into parameterised consecutive quantum gates 
$\theta_1\circ\vec{x},\ldots \theta_L\circ\vec{x}$, where the input data $\vec{x}$ is re-uploaded in every single step before the final result is 
extracted from a final measurement $M$. 
The parameters $\theta_1,\ldots ,\theta_L$ may be optimised classically according to a global cost function in a hybrid classical-quantum algorithm pipeline. 
In contrast to the classical single layered neural network, two major observations can be made: 
\begin{itemize}
    \item[(I)] Despite being applied within a single qubit circuit, the unitary product of all parameterised and re-uploaded gates may introduce highly 
    (up to order $L$) non-linear terms in regard to the input data $\vec{x}$ possibly leading to a much better feature space mapping than accessible by 
    the single layered neural network. 
    \item[(II)] While the classically trivial part of the neural network, i.e.\ the parallel data distribution, needs to be serialised on the qubit register, 
    the classically hard part, i.e.\ the realisation of deep neural networks with many hidden layers, may be much simpler in the quantum circuit. By 
    extending the 1-qubit data re-uploading circuit to $n$ qubits with additional entangling layers in between the re-uploads, one may arrive at a natural 
    augmentation to ``quantum deep learning'' \cite{PerezSalinas2020}.
\end{itemize}

\begin{figure}
    \sidesubfloat[]{\includegraphics[width=0.46\textwidth]{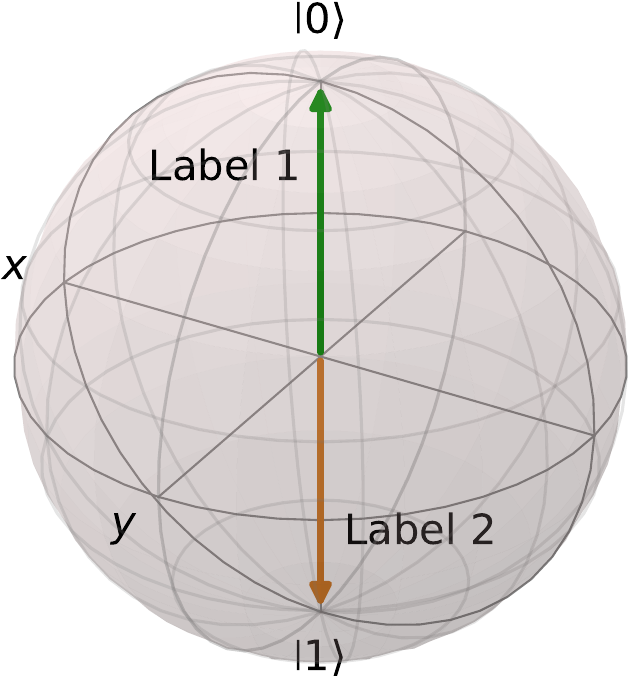}\label{2StateSub.fig}}
    \hfill
    \sidesubfloat[]{\includegraphics[width=0.46\textwidth]{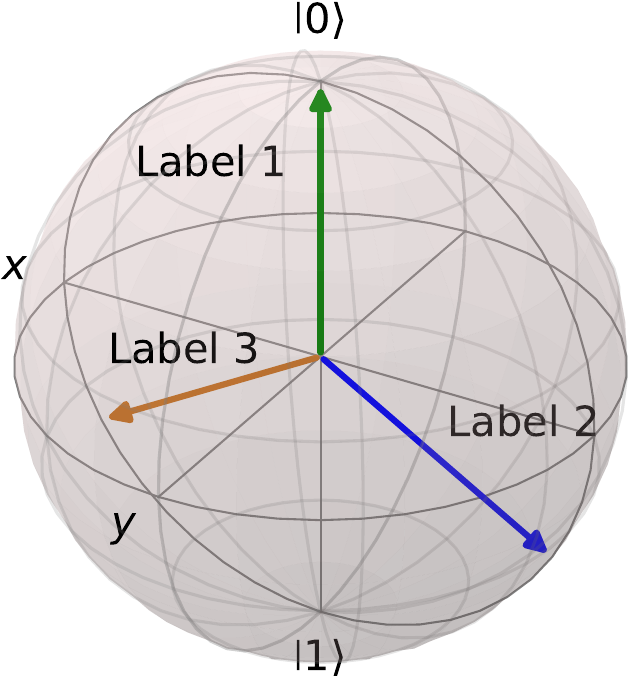}\label{3StateSub.fig}}
    \caption{\label{Label_state_mapping.fig} \textbf{Mapping of classification labels $c$ to maximally orthogonal state vectors $\ket{Y_c}$}. 
    The binary and trinary classification labels $c$ in this work, were mapped to maximally orthogonal label states $Y_c$ on the surface of the single 
    qubit Bloch sphere. This allowed a straightforward association of measured quantum densities to classification labels by means of the respective quantum 
    state fidelity. The label states used in this work are shown in subfigures \ref{2StateSub.fig} and \ref{3StateSub.fig}, respectively.}
\end{figure}

To apply the data re-uploading framework to a multi-label classification task, the measurement outputs need to be mapped to the individual classification 
labels. This can be achieved using maximally orthogonal label states $\ket{Y_c}$ for each classification label 
$c$ as shown in Figure \ref{Label_state_mapping.fig} for binary (subfigure \ref{2StateSub.fig}) and trinary (subfigure \ref{3StateSub.fig}) classification 
tasks.

The classification label $c$ may be acquired by measuring the single qubit state tomography after applying the full data re-uploading circuit 
and identifying the closest mapped label state $\ket{Y_c}$ to the measured density $\tilde{\rho}(\vec{\theta}, \vec{x_{\mu}})$, i.e.\ identifying 
$\ket{Y_c}$ with the largest fidelity to $\tilde{\rho}(\vec{\theta}, \vec{x_{\mu}})$. The circuit parameters may then be trained in a 
supervised learning algorithm to move the single qubit states as close as possible to their respective label states depending on the uploaded input 
data $\vec{x}$. Given that the final classification result is based on a fidelity metric, the UQC pipeline is naturally robust to quantum 
noise \cite{Mccaldin2021}. Given an optimal set of circuit parameters, noise will affect the classified labels 
if and only if the measured density gets closer to another, i.e.\ wrong, maximally orthogonal label state. In other words, binary classification results 
will remain accurate as long as noise does not shift the measured densities to the wrong hemisphere on the Bloch sphere. Analogous reasoning applies to 
multi-label classification problems, where each additional label state decreases the overall noise tolerance.  

For the actual training process, a straightforward cost function $\chi_f^2(\vec{\theta})$ with 
\begin{equation}
    \label{cost.eq}
    \chi_f^2(\vec{\theta}) = \sum_{\mu = 1}^{|T|}\left(1 - \braket{Y_{c(\mu)}|\tilde{\rho}(\vec{\theta}, \vec{x_{\mu}})|Y_{c(\mu)}}\right)
\end{equation}
is based on the quantum state fidelity $\braket{Y_{c(\mu)}|\tilde{\rho}(\vec{\theta}, \vec{x_{\mu}})|Y_{c(\mu)}}$ of the (pure) label state $\ket{Y_{c(\mu)}}$
of the correct label $c(\mu)$ of training data point $\vec{x_{\mu}}$ and the measured density $\tilde{\rho}(\vec{\theta}, \vec{x_{\mu}})$. Clearly, the cost 
function approaches zero if all training data points of the training data set $T$ lead to quantum states coinciding with exact label states.

\section{Methods}
\label{Application}

\subsection{Training process}
\label{Training}

Three different two-dimensional data sets were used in the training process of the UQC model. All sets
contained data points in $\mathbb{R}^2$ within the boundaries of $\left[-1, +1\right]$ and were generated using a random number generator. For each data 
point, labels were generated based on separating margins set by
\begin{itemize}
    \item[(i)] a single circle of radius $\sqrt{\tfrac{2}{\pi}}$ centred at $(0, 0)$, 
    \item[(ii)] the sine function $s(x) = \tfrac{8}{10}\sin (\pi x)$, and 
    \item[(iii)] two circles of radii $1$ and $0.4$ centred at $(-1,-1)$ and $(0.3, 0.3)$, respectively. 
\end{itemize}

For training and validation, two data sets composed of 1,000 and 2,000 data points, respectively, were generated. The training was performed in batches of 100
data points using the Adam optimiser \cite{Kingma2017} employing 20 epochs and a learning rate of $0.6$, as well as $\beta_1 = 0.9$ and $\beta_2 = 0.999$. The trained UQC 
model consisted of parameterised arbitrary rotation gates repeatedly employing unitaries 
\begin{equation}
    U_l = R\left(\vec{v}^{(l)}\right) = R_Z\left(v^{(l)}_3\right)R_Y\left(v^{(l)}_2\right)R_Z\left(v^{(l)}_1\right)
\end{equation}
with 
\begin{equation}
    \vec{v}^{(l)} = 
    \begin{pmatrix}
        v^{(l)}_1 \\
        v^{(l)}_2 \\
        v^{(l)}_3 
    \end{pmatrix}
    =
    \vec{\theta}^{(l)} \circ \vec{x} + \vec{\omega}^{(l)} =  
    \begin{pmatrix}
        \theta^{(l)}_1 x_1 + \omega^{(l)}_1 \\
        \theta^{(l)}_2 x_2 + \omega^{(l)}_2 \\
        \omega^{(l)}_3 \\
    \end{pmatrix}
\end{equation}
for layers $l=1\ldots L$. 
The two-dimensional input data $\vec{x}$ used in this work implies that each layer $U_l$ involves five trainable parameters (two $\theta^{(l)}$ and three $\omega^{(l)}$). 
Six layers were used for the 
binary and ten layers for the trinary classification problems, resulting in 30 and 50 trainable parameters, respectively. During the optimisation, 
the fidelity-based cost function (\ref{cost.eq}) was minimised. All quantum circuits involved in the training process were evaluated classically and off-site 
using an ideal state vector simulator.

\subsection{Deployment to the quantum device}
\label{Deployment}

\begin{figure}
\includegraphics[width=0.96\textwidth, angle=90]{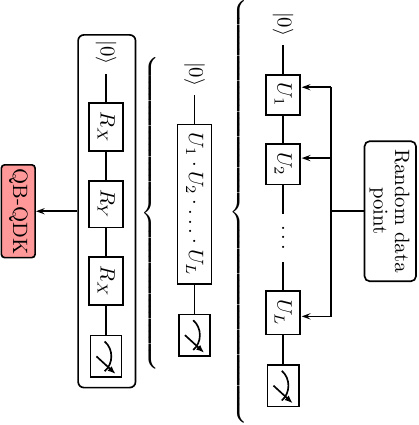}
\caption{\label{QDK_deployment.fig}\textbf{Schematic deployment pipeline of classically optimised UQC models to the QB-QDK quantum computer}. 
To execute inference tasks on the QB-QDK, unseen randomly generated data points were uploaded into the classically optimised UQC models, 
transpiled into a single sequence of $R_XR_YR_X$ rotations and sent off to the room-temperature quantum computer (highlighted in red).}
\end{figure}

After obtaining a trained model in the form of optimised rotation angles $\vec{\theta^{(l)}}$ and $\vec{\omega^{(l)}}$, new random
data points were (i) generated, (ii) inserted into the circuit model, and (iii) classically transpiled into a single sequence of $XYX$ rotations to be 
deployed to our room-temperature quantum computer -- the Quantum Development Kit (QB-QDK). A 
schematic pictogram displaying the deployment route is shown in Figure \ref{QDK_deployment.fig}.

Since the product of two unitaries remains unitary, it is possible to condense all re-uploading layers $1\ldots L$ into a single unitary 
$U_1\cdot U_2\cdot\ldots U_L$, which can then be decomposed (while neglecting an irrelevant global phase) into three single qubit rotations $R_X$, $R_Y$, 
and $R_X$, which are part of the native gate set of the QB-QDK. Note that no approximations are involved, such that
the final condensed circuit still reflects an exact implementation of the classically optimised data re-uploading circuit. Such
simplification techniques are possible due to the limited width of the circuit, but would incur significant overhead in a scaled-up version of the UQC 
(e.g.\ for 30+ qubits). Our results may nevertheless provide a valuable milestone in room-temperature quantum computing because 
\begin{itemize}
    \item[(i)] the deployed models were trained entirely on an ideal state vector simulator, allowing us to directly assess the performance of the QB-QDK 
    against the ideal (noise-free) performance, 
    \item[(ii)] the QB-QDK was operating under real-world conditions in a supercomputing centre, i.e.\ mounted in a HPC server rack 
    experiencing all environmental disturbances such as acoustical and vibrational noise, high temperatures, high air movement, and strong electromagnetic field 
    interference (a detailed performance analysis will be published separately \cite{Akhtar2024}),
    \item[(iii)] no noise mitigation techniques were applied to the measured bit counts.
\end{itemize}

To measure inference of the classically optimised UQC models on the QB-QDK, the condensed $XYX$ circuits were offloaded to the control system of the QB-QDK and 
measured for 100 shots for each data point and measurement basis. In case of the binary classification problems, a single measurement in the computational 
basis was sufficient to estimate the classification label, i.e.\ distinguish between upper and lower Bloch hemispheres (cf.\ Figure \ref{2StateSub.fig}).
Trinary classification problems, required an additional measurement basis orthogonal to the computational basis to estimate the two-dimensional projection 
of the full Bloch vector to the respective label state plane (cf.\ Figure \ref{3StateSub.fig}). 

\subsection{Integration and set up of the quantum device}
\label{Integration}

The QB-QDK is a 19 inch rack-mounted prototype system featuring a room-temperature quantum processing unit (QPU) with 2 qubits built by Quantum Brilliance. 
The quantum system is based on a diamond nitrogen-vacancy (NV) centre and consists of one electronic spin associated with the lattice vacancy and an 
intrinsic $^{14}$N as well as an adjacent $^{13}$C atom \cite{Doherty2013}. The nuclear spin states are used to form the qubit states.

This work has been performed on a unit in the Pawsey Supercomputing Centre server room. For ease of integration, the system's form factor (44.5 cm, 
31 cm and 76 cm for the width, height and depth respectively) complies with standard server rack specifications. An image of the quantum computer in 
the supercomputing centre server room can be seen in Figure \ref{QB_QDK.fig}. The overall power consumption is approximately 180 W and is provided 
by a single mains power cable. The only other external connection is a gigabit fibre connection to the Pawsey networking system, which allows the system 
to receive pre-complied circuits in its native gate set ($R_X$, $R_Y$, C$Z$). The QPU is being kept operational by an automatic calibration system, and is 
interfaced by the full-stack quantum control system developed by Quantum Brilliance. Gate operations are implemented using a series of microwave and 
radio-frequency pulses, that are applied to the NV in the presence of a DC magnetic field \cite{Doherty2013}.

\section{Results}
\label{Results}

\begin{figure}
    \includegraphics[width=0.96\textwidth]{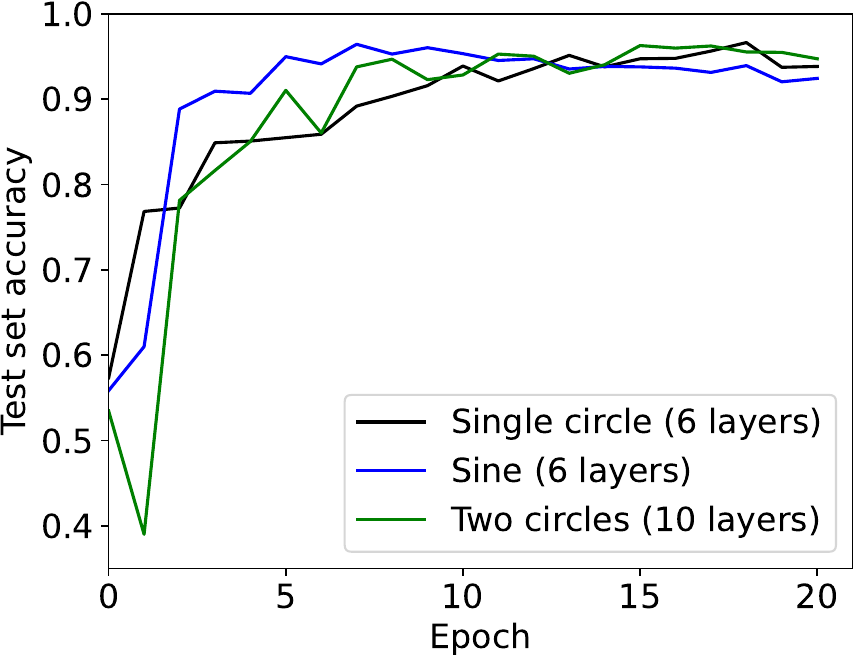}
    \caption{\label{training.fig} \textbf{Training performance of the simulated UQC model}. In every training epoch of each UQC model, their performance was 
    validated against a test data set of 2,000 randomly generated data points. This figure shows the achieved classification accuracy per training epoch and 
    UQC model. For the two binary and the trinary classification problems, six and ten circuit layer repetitions were used, respectively.}
\end{figure}

The classification accuracy for the balanced test data set during the training process is displayed in Figure \ref{training.fig} for the UQC simulated on 
an ideal state vector simulator for the binary circle and sine problems and the trinary two circle problem. 

Although the UQC model was trained using small batches of 100 training data points, its test set performance increased quickly within the first 
few epochs. Independent of the classification problem, optimal test set accuracies above 90\% were reached already within the first ten training epochs. 
There are no qualitative differences between the training performances of the binary (six layers, 30 parameters) and the trinary (ten layers, 50 parameters) 
UQC models. In both cases, the UQC model could be trained efficiently from small training data batches.

\begin{figure*}
    \includegraphics[width=0.32\textwidth]{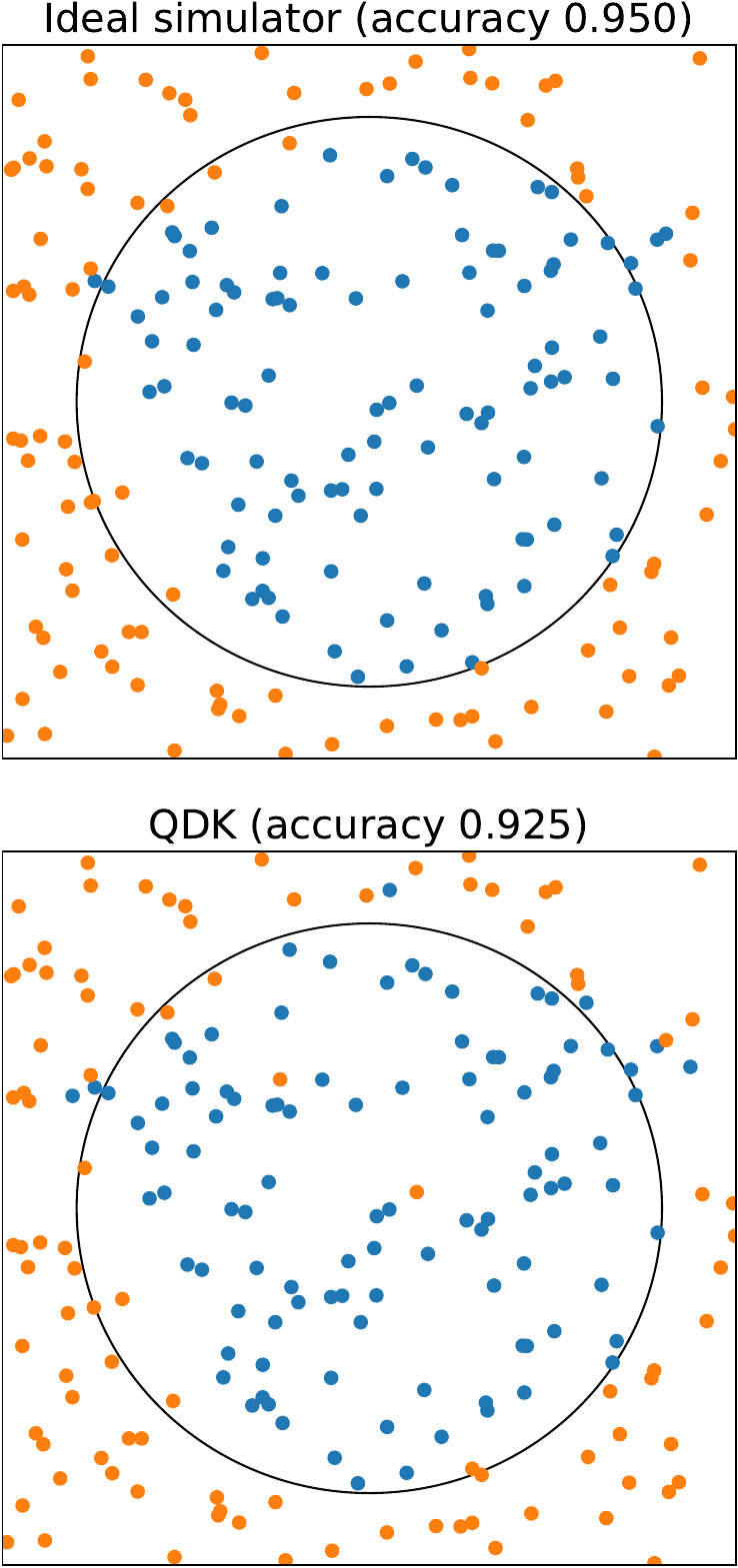}
    \includegraphics[width=0.32\textwidth]{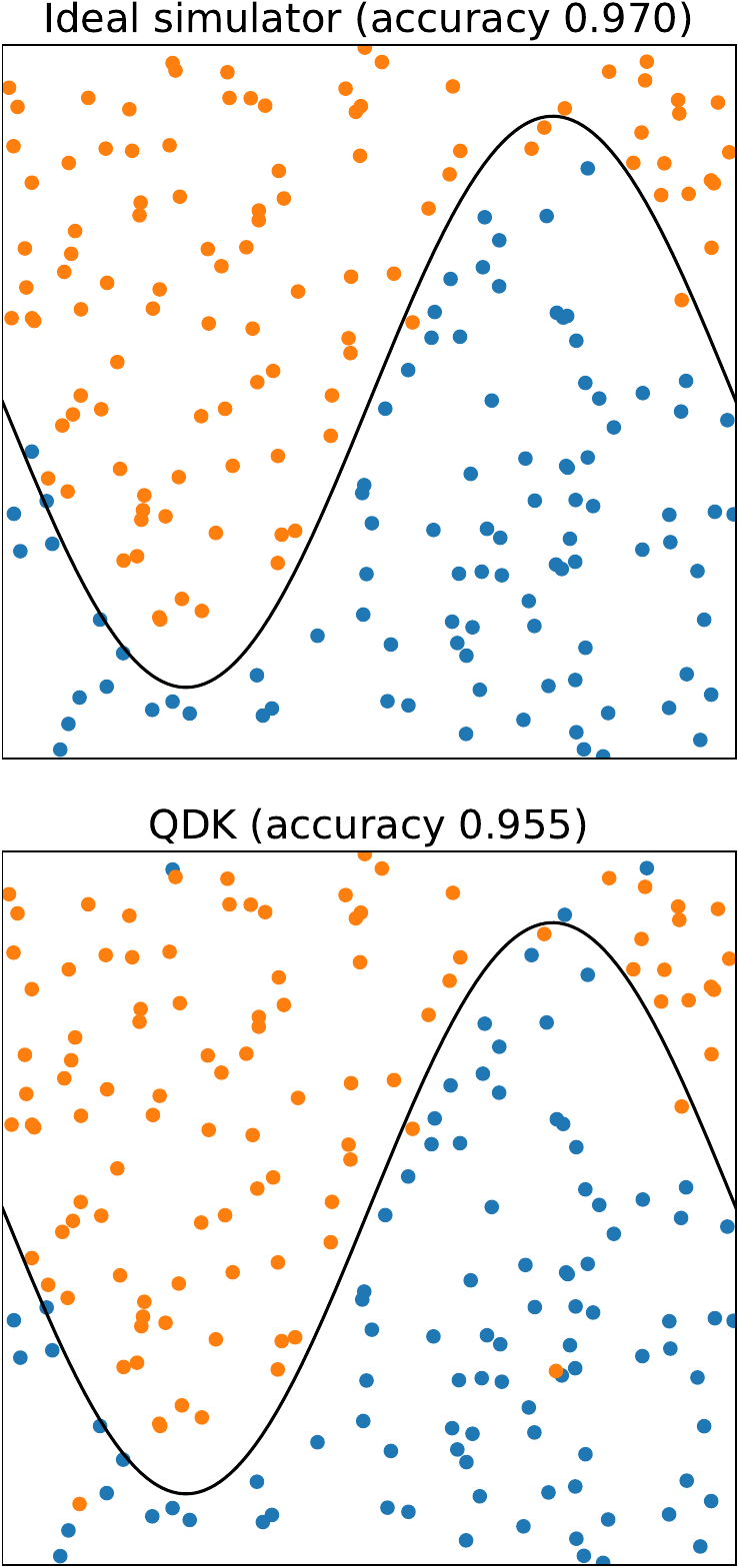}
    \includegraphics[width=0.32\textwidth]{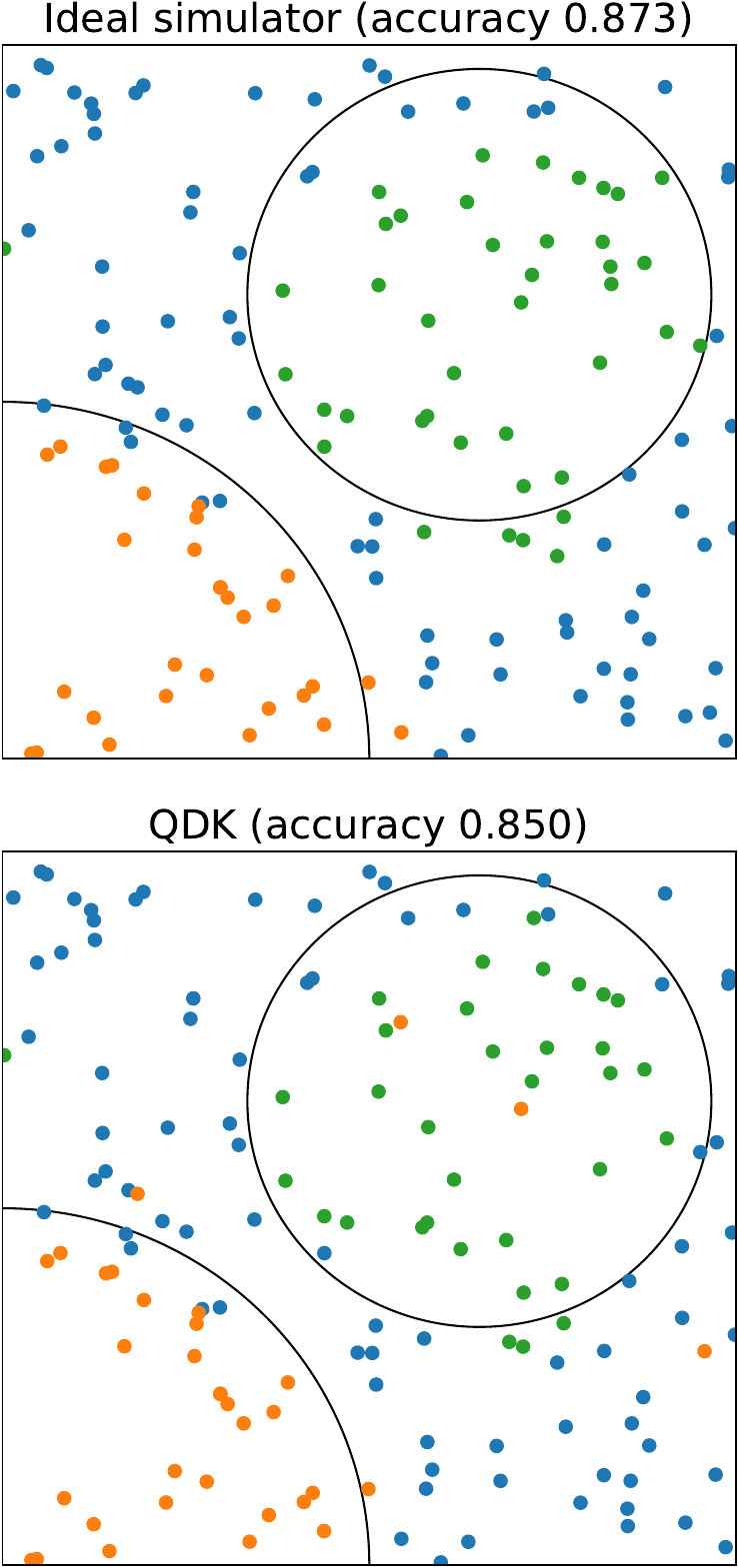}
    \caption{\label{results.fig}\textbf{Classification results of the trained UQC models}. The results for each classification problem obtained from inference 
    of the trained UQC models on an ideal state vector simulator (top) as well as deployed to an on-site room-temperature quantum computer, i.e.\ the QB-QDK 
    (bottom). Each data point is coloured according to the obtained classification label. Separating margins are explicitly shown as a visual aid, only.}
\end{figure*}

Figure \ref{results.fig} shows the classification results of (i) the UQC executed on an ideal simulator, and (ii) the UQC circuits deployed to the 
on-site room-temperature quantum computer QB-QDK.

New, randomly generated data points were used to validate the performance for each classification problem. These are shown as 
individual points in the respective plots. For the binary and trinary classification problems, 200 and 150 data points requiring 20,000 and 30,000 measurements, 
respectively, were evaluated.
The colour of the data points represents the classification result (label) returned by the respective models. The separating margins imposed by the 
training data are included as only a visual aid.

The UQC model could effectively learn all defining features of the training data sets, reaching excellent classification accuracies of 0.95, 0.97, and 
0.873 and 0.925, 0.955, and 0.850 on the ideal state vector simulator and the QB-QDK, respectively. Please note that the optimised classification accuracy 
of the UQC model was above 0.9 for the much larger test data set of the two circle problem (cf.\ Figure \ref{training.fig}). The specific selection of 150 
random data points in Figure \ref{results.fig} lead to a slightly lower performance. In comparison to the ideal state vector results, the QB-QDK performed 
suprisingly well, almost reaching the exact limit of the ideal state vector simulation in terms of the classification accuracy. Overall, the QB-QDK 
reached only 1.5\% to 2.5\% lower classification accuracies. Especially in light of the fact that no on-device training was necessary before deploying 
the UQC circuits to the QB-QDK (cf.\ subsection \ref{Deployment}), these results indicate a significant milestone in room-temperature quantum machine learning 
in the prospect of Quantum Utility \cite{Herrmann2023}. 

Due to quantum noise, a few central data points were seriously misclassified (e.g.\ the two orange data points near the centre of the upper right circle 
in the two circle data set). Such effects could lead to problems in time-constrained QML applications. However, it is reasonable to assume that such 
effects decrease as the performance of the quantum device is optimised. Improved initialisation, gate, and readout fidelities will ultimately lead to 
better classification accuracies, approaching the classical limit set by exact state vector simulations.

\section{Conclusion}

In this work, the first QML applications deployed to an on-site quantum computer operating at ambient conditions (room temperature and ambient 
pressure) -- the QB-QDK -- were reported. The UQC model based on the data re-uploading framework was trained on an ideal state vector simulator. Subsequently, 
the optimised circuits were offloaded to real quantum hardware to infer the labels of newly generated data points. The classification results were compared 
to the exact limit obtainable by ideal state vector simulations.  

The main results collected in Figures \ref{training.fig} and \ref{results.fig} reveal that 
\begin{itemize}
    \item[(i)] the UQC model was able to efficiently learn the defining features of the training data sets, 
    \item[(ii)] the UQC model could be trained in small training data batches, converging to optimal test set classification accuracies within the first 
    10 epochs, 
    \item[(iii)] no on-device training or noise mitigation techniques were required to successfully deploy the classically optimised models to the integrated quantum hardware, 
    reaching classification accuracies only 1.5\% to 2.5\% lower than what is possible with ideal state vector simulators.  
\end{itemize}

Item (iii) marks an important milestone for room-temperature quantum computing, paving a promising path towards Quantum Utility \cite{Herrmann2023} in the 
real world: 
Intermediate sized QML models are trained on large HPC or massively parallel quantum infrastructures and then 
deployed to single on-site quantum accelerators to perform highly complicated inference tasks. Provided 
such tasks are able to outperform similar classical devices in terms of size, weight, and cost in regard to either total runtime, accuracy, or energy 
demands, the first practical application of quantum computers could be realised.

\section*{Acknowledgement}
The authors would like to thank the Pawsey Supercomputing Centre for their support hosting and integrating the first on-premises room-temperature 
quantum computer. In particular, we would like to thank Maciej Cytowski, Marco De La Pierre and Ugo Varetto for the fruitful collaboration and the 
invaluable feedback. The authors would also like to acknowledge funding by the BMBF, in particular projects SPINNING (FKZ 13N16220), QC4DB (FKZ 13N16091),
and QuantumQAP (FKZ: 13N16233).

\bibliographystyle{unsrt}
\providecommand{\noopsort}[1]{}\providecommand{\singleletter}[1]{#1}%

\end{document}